\documentclass{moriond}

\def\be{\begin{equation}}
\def\ee{\end{equation}}
\def\bea{\begin{eqnarray}}
\def\eea{\end{eqnarray}}

\newcommand{\Photo}{}

\begin{document}
\vspace*{4cm}
\title{H.E.S.S. multi-messenger observations}

\author{F. Sch\"ussler on behalf of the H.E.S.S. Collaboration}

\address{IRFU, CEA, Universit\'e Paris-Saclay, F-91191 Gif-sur-Yvette, France}

\maketitle\abstracts{
The H.E.S.S. Imaging Air Cherenkov Telescope system is very well suited to perform follow-up observations of detections at other wavelengths or messengers due to its fast reaction and its comparably low energy threshold. These advantages are fully exploited via a largely automatized system reacting to alerts from various partner observatories covering all wavelengths and astrophysical messengers.\\
In this contribution we provide an overview and recent results from H.E.S.S. programs to follow up on multi-wavelength and multi-messenger alerts. We present ToO observations searching for high-energy gamma-ray emission in coincidence with electromagnetic transients Fast Radio Bursts as well as multi-messenger transients related to high-energy neutrinos and Gravitational Waves.}

\section{High-energy multi-messenger astrophysics}
After decades of ever intensifying studies, the high energy universe is still full of mysteries. A major one is the origin of high-energy cosmic rays. To locate the astrophysical sources and study the acceleration mechanisms able to produce these fundamental particles with energies orders of magnitude above man-made accelerators, is one of the long standing quests in high-energy astrophysics. Thanks to observatories of unprecedented scale and sensitivity, enormous progress has been made over the last years in several aspects. The most significant ones are certainly the wealth of discoveries provided by gamma-ray observatories sensitive in the GeV and TeV energy range, the advent of neutrino astronomy brought about by large scale neutrino telescopes and the direct observation of gravitational waves.

\subsection{The H.E.S.S. high-energy gamma-ray observatory}
\begin{figure}
\begin{minipage}{0.525\linewidth}
\centerline{\includegraphics[width=0.95\linewidth]{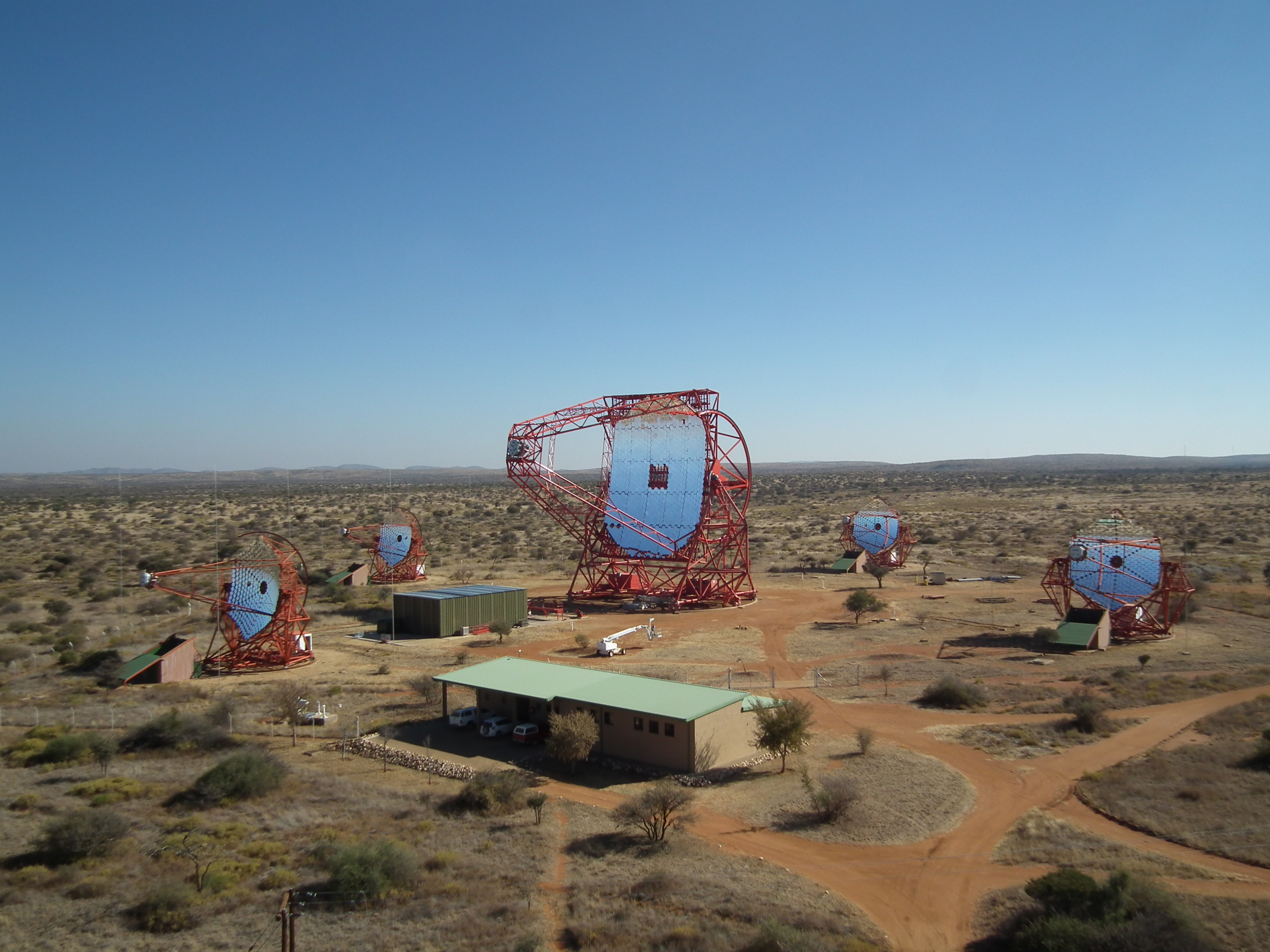}}
\end{minipage}
\hfill
\begin{minipage}{0.47\linewidth}
\centerline{\includegraphics[width=0.95\linewidth]{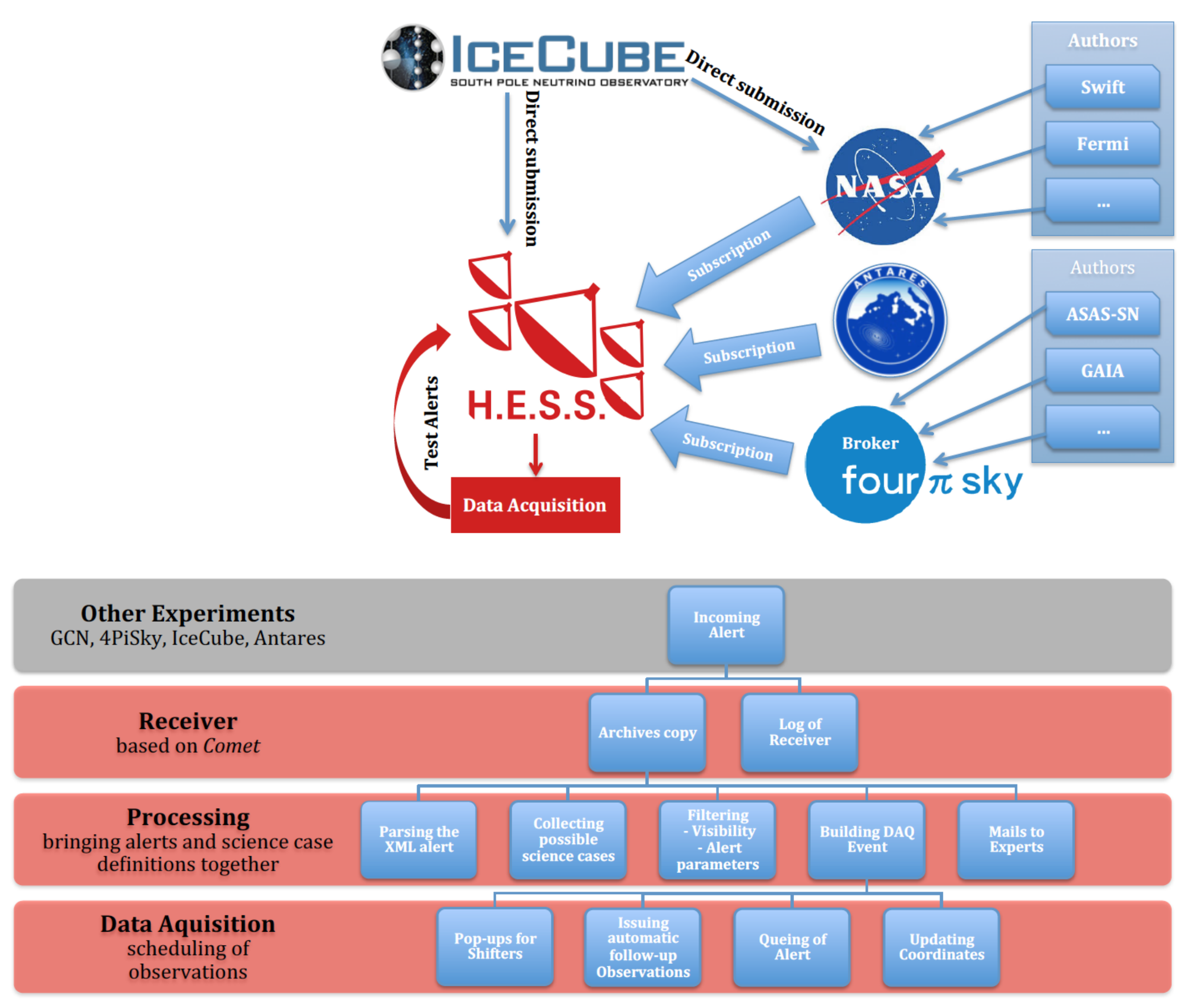}}
\end{minipage}
\caption[]{Left: Photograph of the H.E.S.S. gamma-ray observatory in its final configuration of four 12\,m telescopes combined with a central 28\,m telescope. Right: Schematic view of the H.E.S.S. alert reception system allowing to receive alerts from observatories covering all wavelengths and astrophysical messengers. From~\cite{HoischenBaikal16}.}
\label{fig:HESS}
\end{figure}

The H.E.S.S. imaging atmospheric Cherenkov telescope array is located on the Khomas Highland plateau of Namibia (23$^{\circ}16'18''$ South, $16^{\circ}30'00''$ East), at an elevation of 1800 m above sea level. The original array, inaugurated in 2004, is composed of four telescopes each with a mirror of 12\,m diameter and a camera holding an array of 960 photomultiplier tubes. With this layout H.E.S.S. is sensitive to cosmic and gamma-rays in the 100\,GeV to 100\,TeV energy range and is capable of detecting a Crab-like source close to zenith and under good observational conditions at the 5$\sigma$ level within less than one minute~\cite{HESS-Crab2006}. In 2012 a fifth telescope with a 28\,m diameter mirror was commissioned, extending the covered energy range toward lower energies. The final layout of H.E.S.S. is shown in the left panel of Fig.~\ref{fig:HESS}. 

H.E.S.S. is able to react rapidly to incoming alerts. The installation of the fifth telescope, conceived to allow for extremely rapid slewing reaching every point on the accessible sky within less than one minute~\cite{HESSdrive}, was accompanied by significant efforts to optimize the alert reception and subsequent reaction scheme. The implemented multi-purpose alert system is now connecting the H.E.S.S. observatory to a large variety of observatories worldwide, covering the full wavelength and multi-messenger domain and thus allowing for an extensive multi-messenger program. A schematic view of the alert system is given in the right plot of Fig.~\ref{fig:HESS}, further details can be found in~\cite{HoischenBaikal16}. 

\subsection{High-energy neutrino telescopes}
\begin{figure}
\centerline{\includegraphics[width=0.75\linewidth]{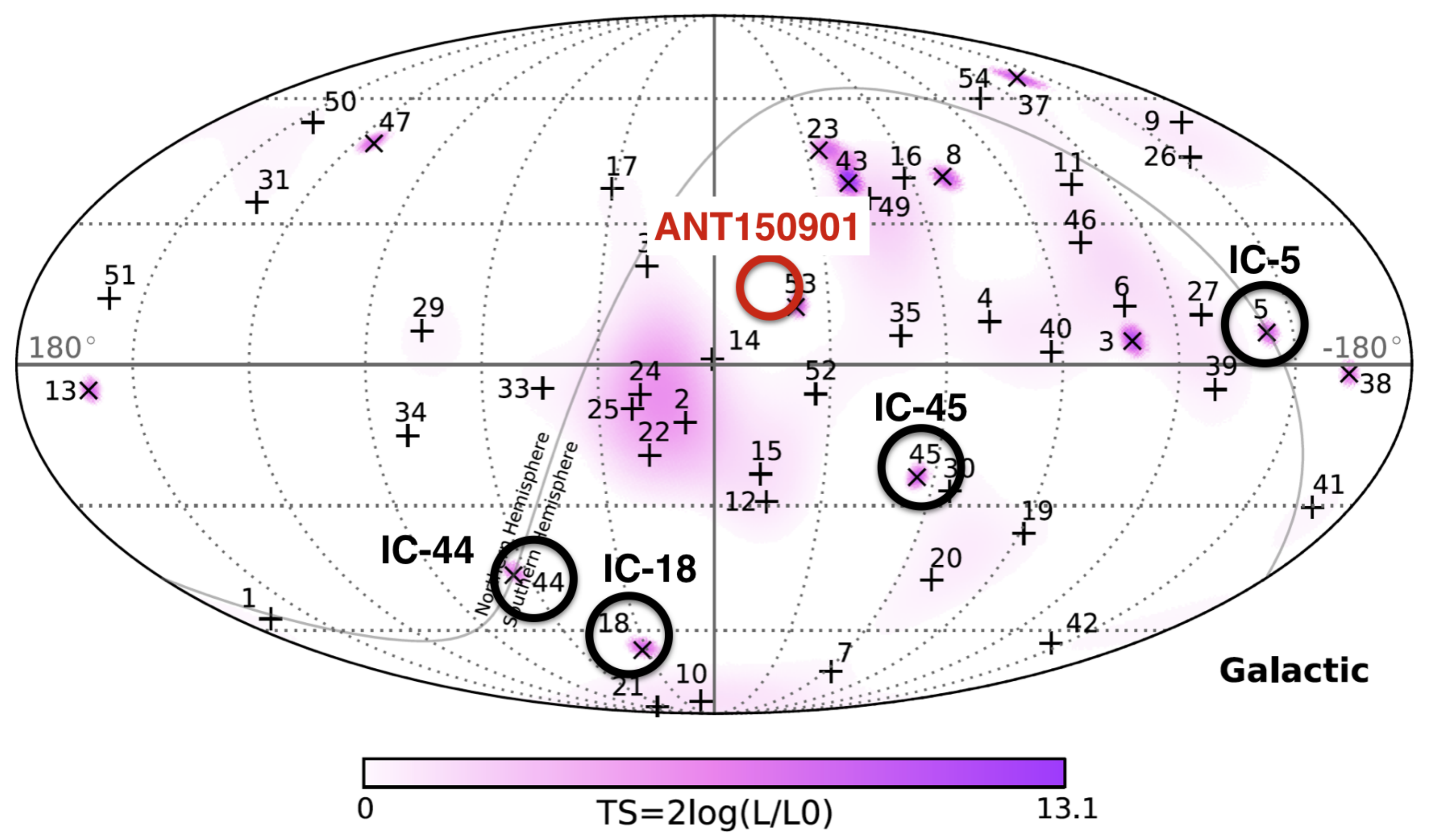}}
\caption[]{The high-energy neutrino sky represented by the arrival directions in Galactic coordinates of events detected by IceCube. We here summarize the H.E.S.S. observations around the events IC-5, IC-18, IC-44 and IC-45 (highlighted by the black circles) and the region of the ANTARES-Swift coincident signal (red circle). Modified from~\cite{IC_HESE_ICRC2015}. }
\label{fig:ICsky}
\end{figure}

Two major instruments searching for astrophysical neutrinos are currently in operation: IceCube~\cite{IceCube} at the South Pole and ANTARES~\cite{ANTARES} in the Mediterranean Sea. So far neither of them has found any significant localized excess~\cite{IC_pointsources}. Yet, a significant breakthrough has been made over the last years by the IceCube collaboration: in four years of data, IceCube was able to single out 54 neutrinos with energies in the range of 60 TeV to 3 PeV that interacted within the instrumented volume (see Fig.~\ref{fig:ICsky}). The atmospheric background contribution has been estimated as $12.6 \pm 5.1$ from cosmic ray muon events and $9^{+8}_{-2.2}$ atmospheric neutrinos events~\cite{IC_HESE_ICRC2015}. However, the origin of these neutrinos is unknown and no significant clustering or excess at small angular scales has been found so far. We here report on searches for high-energy gamma-ray counterparts to these events.

Both neutrino telescopes have implemented systems for rapid event reconstructions, filtering and subsequent alert emission. Whereas the achieved latencies are at the order of several minutes for IceCube~\cite{ICalerts}, the TAToO system of the ANTARES collaboration allows to emit alerts to external observatories within tens of seconds~\cite{TAToO}. The alert reception system of H.E.S.S. is connected to both neutrino telescopes via direct and dedicated links allowing for automatic exchange of information and subsequent follow-up observations. 

\subsection{Gravitational wave observatories}
After substantial upgrades and improvements, the Advanced Ligo interferometers detected the first gravitational waves end of 2015~\cite{LigoDetectionPaper}. This breakthrough is opening a new window to the universe and completes the multi-messenger picture of particles and radiations available to access the high energy univers. 

\begin{figure}
\begin{minipage}{0.525\linewidth}
\centerline{\includegraphics[width=0.9\linewidth]{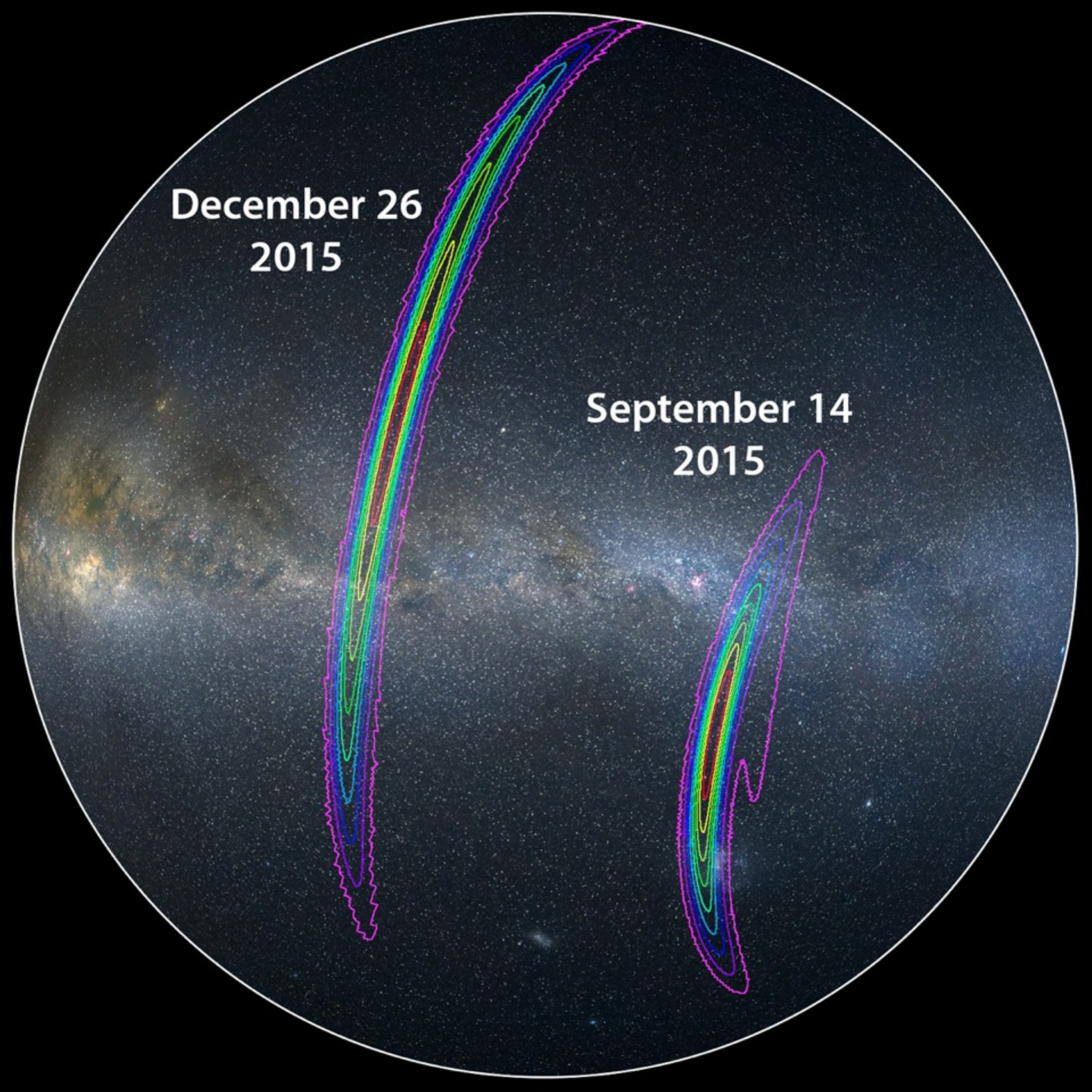}}
\end{minipage}
\hfill
\begin{minipage}{0.46\linewidth}
\centerline{\includegraphics[width=0.9\linewidth]{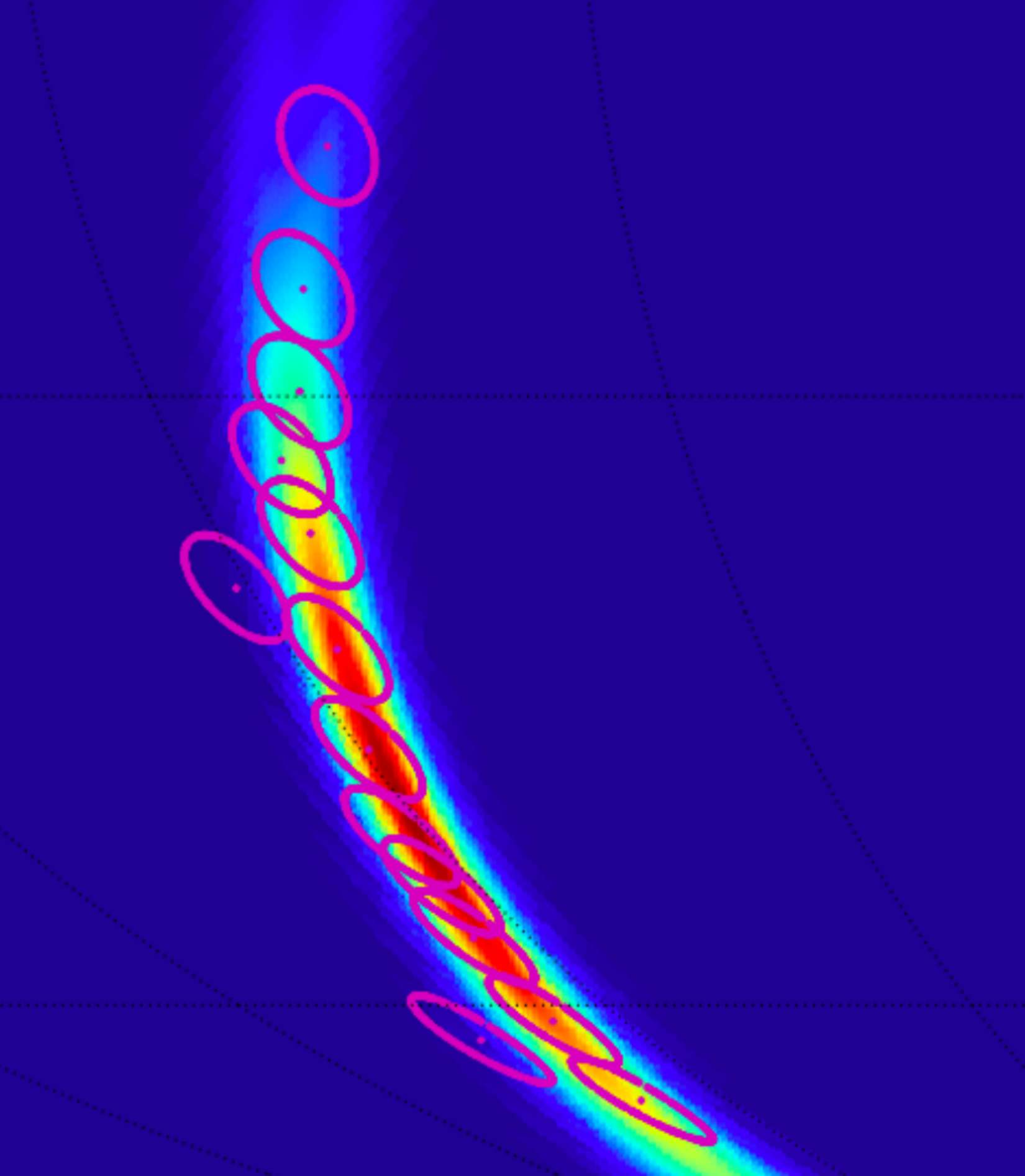}}
\end{minipage}
\caption[]{Left~\cite{IC_HESE_ICRC2015}: Localisation uncertainty of the first two gravitational wave events detected by the Advanced Ligo interferometer. Plot modified from~\cite{LigoDetectionPaper,SkymapViewer}. Right: Potential coverage of H.E.S.S. observations of a simulated gravitational wave event (color scale illustrating the localisation uncertainty) derived with a dedicated, optimized scheduling algorithm. From M. Seglar-Arroyo, these proceedings~\cite{SeglarArroyo_Moriond2017}.}
\label{fig:Ligo}
\end{figure}

The localisation of the massive binary black hole systems whose mergers caused the emission of the observed gravitational waves was unfortunately rather imprecise (cf. left plot of Fig.~\ref{fig:Ligo}). Nevertheless, a large number of observatories covering all wavelengths from radio to gamma rays are participating in a global effort to search for electromagnetic (and high-energy neutrino) counterparts to the gravitational wave events. So far, no clear detection could be made. The H.E.S.S. experiment is participating in this effort. Due to the large localisation uncertainties (hundreds to thousands of square degrees) dedicated follow-up strategies have to be put into place. An outline of current strategies, implemented within the general, automatic alert reception system mentioned above, is given in a dedicated contribution to this conference~\cite{SeglarArroyo_Moriond2017}. An illustration of the performance of the coverage achieved for a simulated gravitational wave event is shown in the right plot of Fig.~\ref{fig:Ligo}. For this exemplary event we cover most of the uncertainty area accessible from the location of the H.E.S.S. experiment within only a few hours (14 individual pointings). With the default duration of a H.E.S.S. data taking run of 30\,min, this leads to a sensitivity to fluxes of about 5\% of the flux from the Crab nebula throughout the region defined by the gravitational wave uncertainty map.

\section{H.E.S.S. multi-messenger searches}
The main focus of the H.E.S.S. multi-messenger program is to exploit the intimate connection between high-energy neutrinos and gamma-rays. Provided appropriate conditions of the environment of cosmic accelerators (e.g. magnetic fields, matter and field densities, etc.), high-energy (hadronic) particles are potentially undergoing interactions with matter and radiations fields within and/or surrounding the acceleration sites. The light mesons, predominantly pions, created in these interactions will decay by emitting both high-energy neutrino as well as gamma-rays. For sources where the matter and radiation fields are not too dense to cause absorption of the emitted gamma-rays, we can therefore hope to find spatial and temporal correlated emission of both messengers. 

The combination of observations of different messengers also allows to increase the sensitivity of the searches. This is because every messenger has its own background which has to be overcome in order to detect a significant signal and/or allow to confidently interpret the measurements. In the search for the astrophysical sources of high-energy cosmic rays, these backgrounds are both astrophysical as well terrestrial. Thanks to their high sensitivity, gamma-ray observatories operating in the GeV-TeV energy range detected a multitude of emission regions over the last years. With the exception of the discovery of a pion decay signature in low energy gamma rays from two supernova remnants by Fermi~\cite{FermiPionBump} and the discovery of PeV proton acceleration near the Galactic Center~\cite{Pevatron}, this is usually not sufficient proof for the presence of accelerated hadronic CRs as gamma-ray radiation can also be induced by accelerated electrons (via Bremsstrahlung or inverse Compton scattering of low energy photons). For most detected gamma-ray sources, both leptonic (accelerating mainly electrons) as well as hadronic (accelerating predominantly hadronic CRs) models are able to explain the observed emission. Attempts to distinguish between these competing explanations are usually based on the spectral shape of the GeV-TeV emission and a clear discrimination between different models is not possible. 

High-energy neutrino telescopes suffer from different challenges: their effective areas are orders of magnitude below current gamma-ray observatories and available event statistics are relatively low. In addition, high-energy neutrinos are produced copiously in the Earth's atmosphere via CR induced extensive air showers. These {\it atmospheric neutrinos} are an important background for the search of astrophysical neutrino sources but their influence can be reduced thanks to their soft energy spectrum following $E^{-3.6}$ (compared to the harder $E^{-2}$ spectrum expected from Fermi acceleration processes in astrophysical sources). On the other hand, high-energy neutrinos are a clear sign of hadronic processes and, in contrast to most gamma-ray observatories, neutrino telescopes have the capability to observe large portions of the sky without the necessity of scheduled, pointed observations. They are therefore ideally suited to monitor the high-energy universe and provide alerts on interesting events for detailed follow-up observations by other instruments.


Searches for coincident emission of both, high-energy neutrinos and gamma-rays overcomes most of the mentioned problems and may therefore allow to locate the long sought acceleration sites of high-energy cosmic rays. We present some of these searches performed within the H.E.S.S. multi-messenger program in the following.

\subsection{H.E.S.S. observations of regions around IceCube high-energy neutrino events} 
Over the last years we started to exploit these potential correlations in searches for high energy gamma-ray emission from regions surrounding the arrival direction of high energy neutrinos detected by IceCube. Over several years, four of the published events fulfilling the "High-energy Starting Event" (HESE)~\cite{IC_HESE_ICRC2015} criteria have been scanned with the H.E.S.S. array. A summary of the events and the obtained observations is given in Table~\ref{tab:HESE}. The data described here have been taken with the full array of all five H.E.S.S. telescopes. During the analysis we require that data from at least  two telescopes participated in the reconstruction of the gamma-ray induced air shower. The data were reconstructed using the Model Analysis~\cite{ModelAnalysis}, an advanced Cherenkov image reconstruction method in which the recorded shower images of all triggered telescopes are compared to a semi-analytical model of gamma-ray showers by means of a log-likelihood optimization. Reconstructed events have to fulfill a {\it loose} cut configuration allowing for a comparably low energy threshold. 

\begin{figure*}[!t]
\centering
\includegraphics[width=0.85\textwidth]{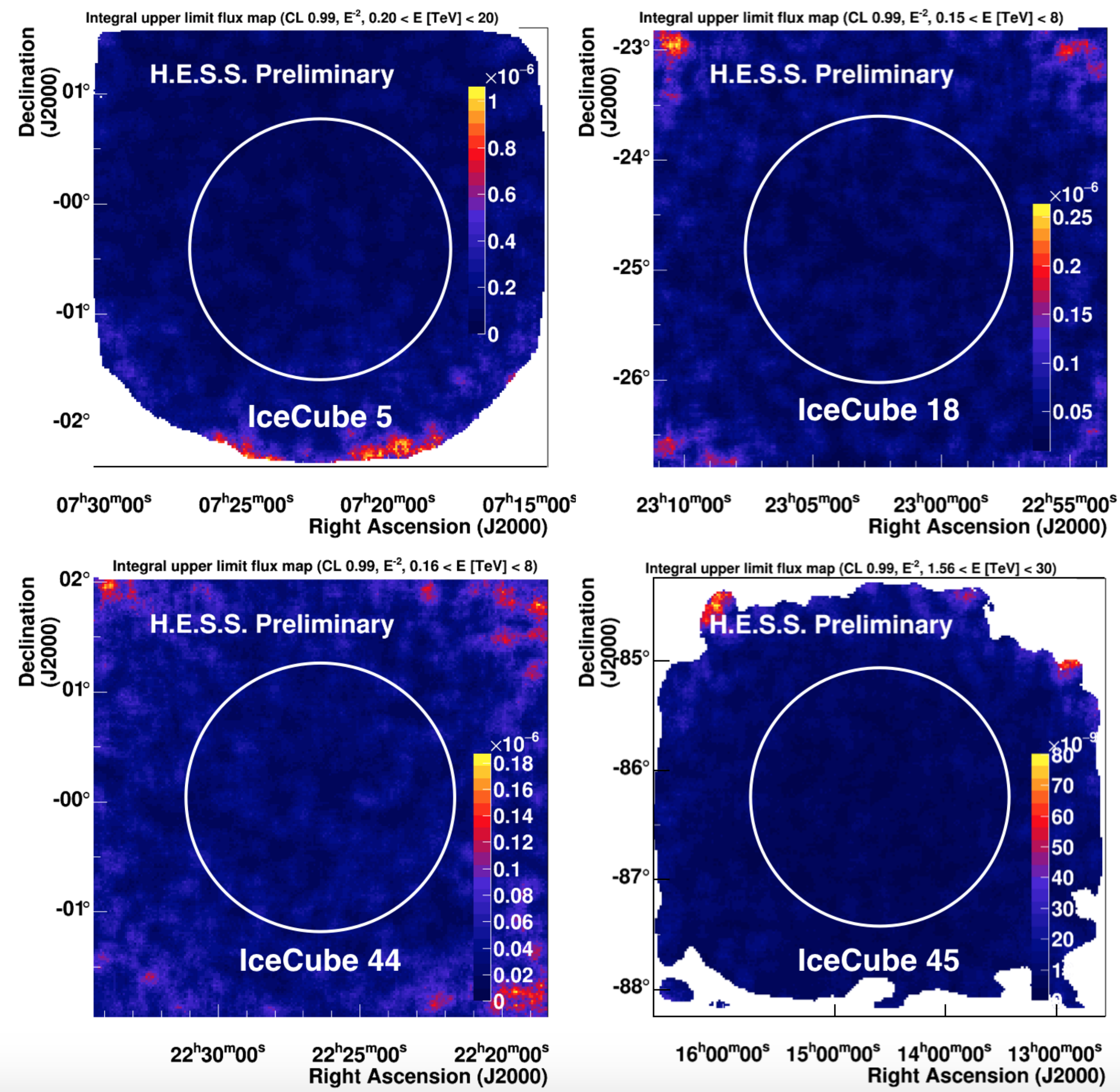}
\caption{Skymaps of regions around high-energy neutrinos recorded by IceCube. The color scale represents upper limits in units of $\mathrm{m}^{-2} \mathrm{s}^{-1}$ on the high-energy gamma-ray flux derived from H.E.S.S. observations. The white circles denote the localisation uncertainty of the neutrino events (cf. Tab.~\ref{tab:HESE}).}\label{fig:HESSUL}
\end{figure*}

\begin{table*}[th]
\caption{High-energy neutrino events used for H.E.S.S. searches}
\label{tab:HESE}
\centering
\begin{tabular}{llllll}
\hline
&IC-5 & IC-18 & IC-44& IC-45   \\
\hline
 {\bf IceCube}& & &  &\\
Right Ascension [h]& 7.37 & 23.04 & 22.4 & 14.59 \\
Declination [$^\circ$]& -0.4 & -24.8 & 0.04 & -86.25 \\
 Median angular resolution [$^{\circ}$]& $<$ 1.2 & $<$ 1.3 & $<$ 1.2 & $<$ 1.2  \\
 Deposited energy [TeV]& $71.4_{-9.0}^{+9.0}$ & $31.5_{-3.3}^{+4.6}$ & $84.6_{-7.9}^{+7.4}$ & $429.9^{+57.4}_{-49.1}$  \\
 \hline
{\bf H.E.S.S. observations}& & &  &\\
 Zenith angle range [$^{\circ}$]& 25 - 35 & 2 - 20 & 25 - 40 & 60 - 66  \\
 Effective observation time [min] & 72 & 486 & 432 & 270  \\
\hline
 \end{tabular}
 \end{table*}

None of the observed regions showed significant high-energy gamma-ray emission. To derive upper limits on the gamma-ray flux we have to take into account the large positional uncertainty related to the neutrino events, which is of the order of $1.2\,\mathrm{deg}$ (see Tab.~\ref{tab:HESE} for details). We here derive maps showing integral flux upper limits for the whole region-of-interest. The limits are derived above an energy threshold $E_\mathrm{thr}$ defined individually for each observation as the energy where the acceptance is $10\%$ of its maximum value and yielding more than 10 events available to estimate the background. The background level in the field-of-view was determined from the dataset itself using the standard ``ring background'' technique~\cite{RingBg}. In order to build the upper limits maps we follow the general idea as developed for the  H.E.S.S. Galactic Plane Survey~\cite{HGPS_Gamma16}. The computation is following
\begin{equation}
F = \frac{N^{\mathrm{UL}}_\gamma}{N_\mathrm{exp}} \int_{E_\mathrm{min}}^{E_\mathrm{max}} \Phi_\mathrm{ref}(E) \; \mathrm{d}E,
\end{equation}
where $F$ is the integral flux computed for each spatial bin of the map between $E_\mathrm{min}$ and $E_\mathrm{max}$. Here we set $E_\mathrm{min}= E_\mathrm{thr}$ to the energy threshold of each observation (cf. Tab.\ref{tab:HESE}).  $N^{\mathrm{UL}}_\gamma$ is the upper limit on the number of gamma-ray events in each bin of the map. It is obtained for a confidence level of $99\%$~\cite{Rolke}. $N_\mathrm{exp}$ is the total predicted number of events. It is given by 
\begin{equation}
N_\mathrm{exp} = \sum_{i=0}^{i=N_\mathrm{runs}} T_i \int_{E_\mathrm{thr}}^{\infty} \Phi_\mathrm{ref}(E_\mathrm{rec}) A_\mathrm{eff}(E_\mathrm{rec}, R_i) \; \mathrm{d}E_\mathrm{rec} .
\end{equation}
Here, $E_\mathrm{rec}$ is the reconstructed energy, $T_i$ is the effective live time and $R_i$ symbolizes the observation parameters for run $i$ (zenith, off-axis and azimuth angle, pattern of telescopes participating in the run, optical efficiencies). $A_\mathrm{eff}$ is the effective area and $E_\mathrm{thr}$ is the threshold energy appropriate for the observation. The function $\Phi(E)$ is the assumed gamma-ray spectral energy distribution. Here we use a generic power law following $E^{-2}$. 

The resulting upper limit skymaps of the regions around the four selected IceCube HESE events are shown in Fig.~\ref{fig:HESSUL}.

\section{Multimessenger searches for transient phenomena}
Most astrophysical sources emitting in the high-energy domain show transient behavior, i.e. the emitted flux is highly variable. Due to the joint emission mechanism outlined above, we can expect close correlations in both the spatial and the time domain between different messengers. If found, coincident detections of high-energy neutrinos and gamma-rays would provide the smoking gun for hadronic interaction processes and thus provide important hints towards the long sought cosmic ray sources. It should be noted that gamma-ray induced pair creation in the potentially dense radiation fields of sources might dilute the temporal coincidences between the emission of TeV gamma-rays and TeV-PeV neutrinos. Due to the largely unknown source properties, a large discovery space remains nevertheless uncharted.

\subsection{H.E.S.S. observations of the joint ANTARES-Swift transient ANT150901A} 
A first H.E.S.S. search for transient sources using a multi-messenger approach was performed in September 2015 with the follow-up of the ANTARES neutrino alert ANT150901A. After the detection of a high-energy neutrino by the online reconstruction of the ANTARES neutrino telescope on September 1st, 2015, at 07:38:25 UT, an alert has been issued to a variety of optical telescopes and the Swift X-ray satellite~\cite{TAToO}. The region identified by the neutrino direction (RA=$246.43\,\mathrm{deg}$, Dec=$-27.39\,\mathrm{deg}$ with an of uncertainty radius of $18\,\mathrm{arcmin}$) has been observed 10 hours later by Swift. An unknown, relatively bright ($\Phi = 5\times10^{-13} - 1.4\times10^{-12}\,\mathrm{erg}\,\mathrm{cm}^{-2}\,\mathrm{s}^{-1}$ at $0.3 - 10\,\mathrm{keV}$) and variable X-ray source has been detected within the neutrino error circle. These observations where reported in ATEL\#7987

\begin{figure}
\begin{minipage}{0.4\linewidth}
\centerline{\includegraphics[width=0.95\linewidth]{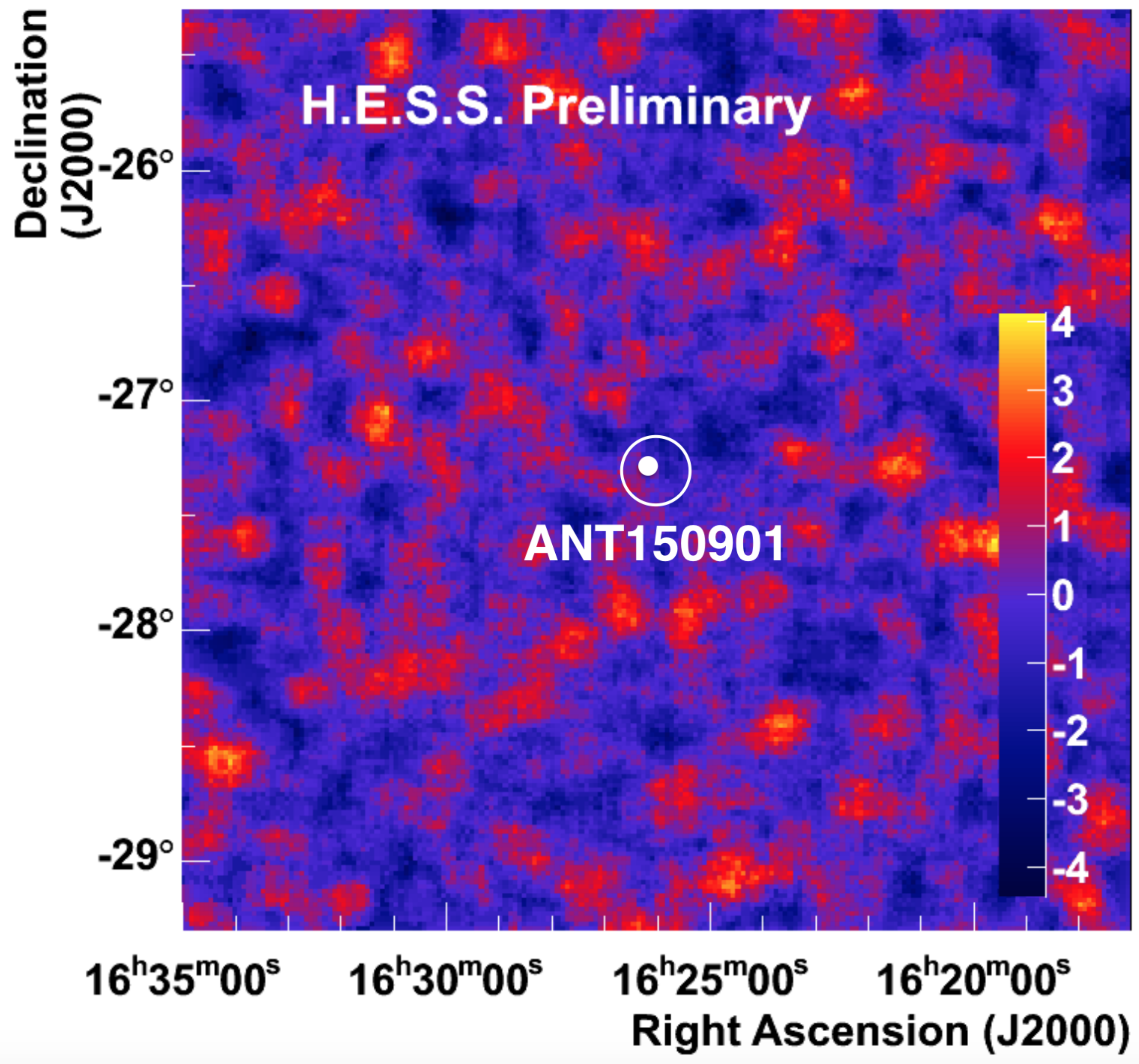}}
\end{minipage}
\hfill
\begin{minipage}{0.58\linewidth}
\vspace{-5mm}
\centerline{\includegraphics[width=0.95\linewidth]{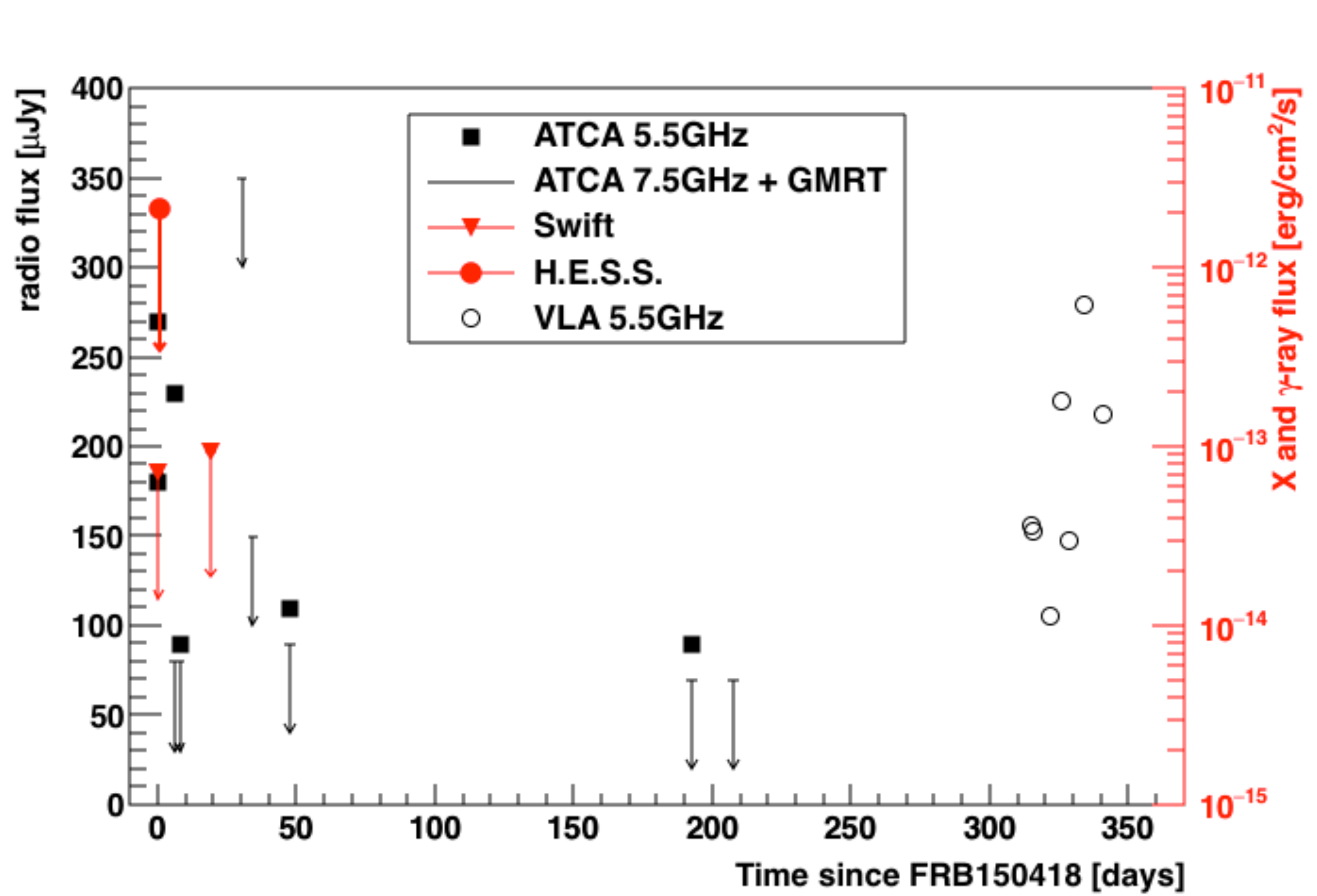}}
\end{minipage}
\caption[]{Left: Significance map derived from H.E.S.S. follow-up of the region around the ANTARES high-energy neutrino alert ANT150901A published in ATEL\#7987. The uncertainty of the neutrino direction ($0.3\,\mathrm{deg}$) is shown as white circle and the location of the detected variable Swift source is indicated by the white marker. Right: Summary of the multi-wavelength observations following the detection of FRB150418 by SUPERB@Parkes illustrating the discovery of a radio afterglow by ATCA (black squares). The upper limits obtained from H.E.S.S. observations are shown as red marker. The black open circles show the renewed radio activity from the region found in VLA observations.}
\label{fig:ANT150901A}
\end{figure}

H.E.S.S. follow-up observations have been scheduled immediately. They started September 3rd, 2015, at 18:58 UT as soon as the necessary observation conditions were reached. The significance map derived from $1.5\,\mathrm{h}$ of observations is shown in Fig.~\ref{fig:ANT150901A}. The uncertainty on the direction of the high-energy neutrino is shown as white circle and the location of the Swift source is indicated by the white marker. Without the detection of high-energy gamma-ray emission we derived an upper limit on the gamma-ray flux to $\Phi(\mathrm{E}>320\,\mathrm{GeV}, 99\,\%\,\mathrm{C.L.}) < 2.7 \times 10^{-8}\,\mathrm{m}^{-2}\,\mathrm{s}^{-1}$. It should be noted that the extensive multi-wavelength follow-up of ATEL\#7987 lead finally to the conclusion that the Swift X-ray source is due to a young and/or variable star (USNO-B1.0 0626-0501169) and thus unrelated to the neutrino, which may be of atmospheric origin.

\subsection{First limits on the gamma-ray afterglow of a Fast Radio Burst} 
Fast Radio Bursts (FRBs) are a relatively new class of very brief, yet powerful transient phenomena observed in the radio domain. A summary of known FRBs including the details of the observations can be found in the online catalogue FRBcat~\cite{FRBcat}. H.E.S.S. is participating in the search for FRB counterparts via a collaboration with the SUPERB project operating at the Parkes telescope.

A particular interesting FRB has been detected on 18th April 2015 by the SUPERB team at the Parkes telescopes. Following the detection, the ATCA array was able to detect a radio afterglow signal from the direction of the burst~\cite{FRB150418_Keane} lasting up to 6 days (cf. Fig.~\ref{fig:ANT150901A}). Optical follow-up observations allowed to link it to an elliptical galaxy at $z=0.492\pm0.008$  (WISE\,J071634.59$-$190039.2), which, if confirmed, would be the first identified FRB host galaxy. Further observations of the region have been obtained after the publication of the initial results with a variety of instruments. A selection of them is shown in Fig.~\ref{fig:ANT150901A}. As illustrated, renewed radio activity from the region has been found almost a year after the burst and thus raising doubts on the connection between the FRB and the radio afterglow.

H.E.S.S. follow-up observations could be obtained about 14.5\,h after the FRB as soon as the necessary darkness conditions were fulfilled. In 1.4\,h of data, no significant gamma-ray afterglow emission could be found. Integrating above the threshold of $350\,\mathrm{GeV}$ we derived an upper limit on the gamma-ray flux to $\Phi_\gamma(E > 350\,\mathrm{GeV}, 99\,\%\,\mathrm{C.L.}) < 1.33\times 10^{-8}\,\mathrm{m}^{-2} \mathrm{s}^{-1}$, assuming a generic $E^{-2}$ energy spectrum. It is shown in comparison to other observations in the right plot of Fig.~\ref{fig:ANT150901A} and represents the first limit on the gamma-ray afterglow of fast radio bursts~\cite{HESSFRBpaper}.

\section{Summary and outlook}
In this contribution we presented an updated status of the H.E.S.S. multi-messenger program searching for coincidences of high-energy gamma rays with high-energy neutrinos and gravitational waves but also with novel transient phenomena like Fast Radio Bursts. Its rapid reaction combined with an automatic and versatile alert reception system and its unprecedented sensitivity makes H.E.S.S. the prime observatory for these searches. Hopes are thus high that the detection of the first astrophysical high-energy multi-messenger source may be just around the corner.

\section*{Acknowledgments}
{\scriptsize The support of the Namibian authorities and of the University of Namibia in facilitating the construction and operation of H.E.S.S. is gratefully acknowledged, as is the support by the German Ministry for Education and Research (BMBF), the Max Planck Society, the German Research Foundation (DFG), the Alexander von Humboldt Foundation, the Deutsche Forschungsgemeinschaft, the French Ministry for Research, the CNRS-IN2P3 and the Astroparticle Interdisciplinary Programme of the CNRS, the U.K. Science and Technology Facilities Council (STFC), the IPNP of the Charles University, the Czech Science Foundation, the Polish National Science Centre, the South African Department of Science and Technology and National Research Foundation, the University of Namibia, the National Commission on Research, Science \& Technology of Namibia (NCRST), the Innsbruck University, the Austrian Science Fund (FWF), and the Austrian Federal Ministry for Science, Research and Economy, the University of Adelaide and the Australian Research Council, the Japan Society for the Promotion of Science and by the University of Amsterdam.
We appreciate the excellent work of the technical support staff in Berlin, Durham, Hamburg, Heidelberg, Palaiseau, Paris, Saclay and in Namibia in the construction and operation of the equipment. This work benefited from services provided by the H.E.S.S. Virtual Organisation, supported by the national resource providers of the EGI Federation.}


\end{document}